\documentclass[journal,10pt]{IEEEtran}
\usepackage{cite}
\usepackage{amsmath}
\usepackage{breqn}
\usepackage{mathrsfs}
\usepackage[mathscr]{euscript}
\usepackage{algorithm}
\usepackage{xspace}
\usepackage{algorithmic}
\usepackage[dvipsnames]{xcolor}
\usepackage{tabularx}
\usepackage{array}
\usepackage{stfloats}
\usepackage{graphicx}
\usepackage{subcaption}
\usepackage[justification=justified]{caption}
\usepackage{optidef}
\usepackage{amsmath}
\usepackage{amssymb}
\usepackage{amsthm}
\usepackage{lipsum}
\usepackage[breaklinks]{hyperref}
\usepackage{microtype}
\usepackage{bbm}
\newcolumntype{C}[1]{>{\centering\arraybackslash}p{#1}}
\newtheorem{theorem}{Theorem}
\newtheorem{definition}{Definition}

\newcommand{\eg}{\textit{e.g.,}\xspace}
\newcommand{\ie}{\textit{i.e.,}\xspace}

\begin{document}
\title{Dynamic SDN-based Radio Access Network Slicing with Deep Reinforcement Learning for URLLC and eMBB Services}

\author{
    \IEEEauthorblockN{
    	Abderrahime~Filali,~\IEEEmembership{Member,~IEEE,}
    	Zoubeir~Mlika,~\IEEEmembership{Member,~IEEE,}
    	Soumaya~Cherkaoui,~\IEEEmembership{Senior Member,~IEEE,}
    	and~Abdellatif~Kobbane,~\IEEEmembership{Senior Member,~IEEE}
    }

    \thanks{
		A. Filali, Z. Mlika, and S. Cherkaoui are with the INTERLAB Research Laboratory, Faculty of Engineering, Department of Electrical and Computer Science Engineering, Université de Sherbrooke, Sherbrooke (QC) J1K 2R1, Canada 
		(e-mails: abderrahime.filali@usherbrooke.ca, zoubeir.mlika@usherbrooke.ca, soumaya.cherkaoui@usherbrooke.ca).

		A. Kobbane is with the UM5R ENSIAS, BP 713 Rabat, Morocco (e-mail: abdellatif.kobbane@um5.ac.ma).
	}
}

\maketitle

\begin{abstract}
Radio access network (RAN) slicing is a key technology that enables 5G network to support heterogeneous requirements of generic services, namely ultra-reliable low-latency communication (URLLC) and enhanced mobile broadband (eMBB). In this paper, we propose a two time-scales RAN slicing mechanism to optimize the performance of URLLC and eMBB services. In a large time-scale, an SDN controller allocates radio resources to gNodeBs according to the requirements of the eMBB and URLLC services. In a short time-scale, each gNodeB allocates its available resources to its end-users and requests, if needed, additional resources from adjacent gNodeBs. We formulate this problem as a non-linear binary program and prove its NP-hardness. Next, for each time-scale, we model the problem as a Markov decision process (MDP), where the large-time scale is modeled as a single agent MDP whereas the shorter time-scale is modeled as a multi-agent MDP. We leverage the exponential-weight algorithm for exploration and exploitation (EXP3) to solve the single-agent MDP of the large time-scale MDP and the multi-agent deep Q-learning (DQL) algorithm to solve the multi-agent MDP of the short time-scale resource allocation. Extensive simulations show that our approach is efficient under different network parameters configuration and it outperforms recent benchmark solutions.

\end{abstract}

\begin{IEEEkeywords}
Network slicing, software defined networking, URLLC, eMBB, deep reinforcement learning. 
\end{IEEEkeywords}

\section{Introduction}
The heterogeneous services supported by the fifth-generation (5G) new radio (NR) can be classified mainly into enhanced mobile broadband (eMBB), ultra-reliable low-latency communication (URLLC) and massive machine-type communication (mMTC) services \cite{survey5G}. The eMBB services target the applications that require a high data rate such as high definition (HD) video or large-scale video streaming. The URLLC services accommodate low-latency and high reliability applications such as autonomous driving or robotic surgery. Finally, the mMTC services provide connectivity to a large number of devices, e.g., massive access in Internet of things (IoT) networks~\cite{z_iot,nour_iot}, that are characterized by small data and sporadic traffic. To support these three 5G services while respecting their heterogeneous and different requirements over a common wireless network infrastructure, radio access network (RAN) slicing \cite{ranSlicing} is introduced as a key enabling technology in the new generation of cellular networks. Network slicing (NS) provides the ability to build several independent logical networks, called network slices, each adapted to the requirements of a specific service \cite{NS_MEC}. Therefore, each RAN slice can be tailored and dedicated to support a specific service with distinctive characteristics and requirements. The network operator can leverage the network programmability provided by software defined networking (SDN) to dynamically manage the provisioning of radio resources for the RAN slices \cite{SDN_NS1}. 
\begin{figure}[t]
	\includegraphics[width=\linewidth]{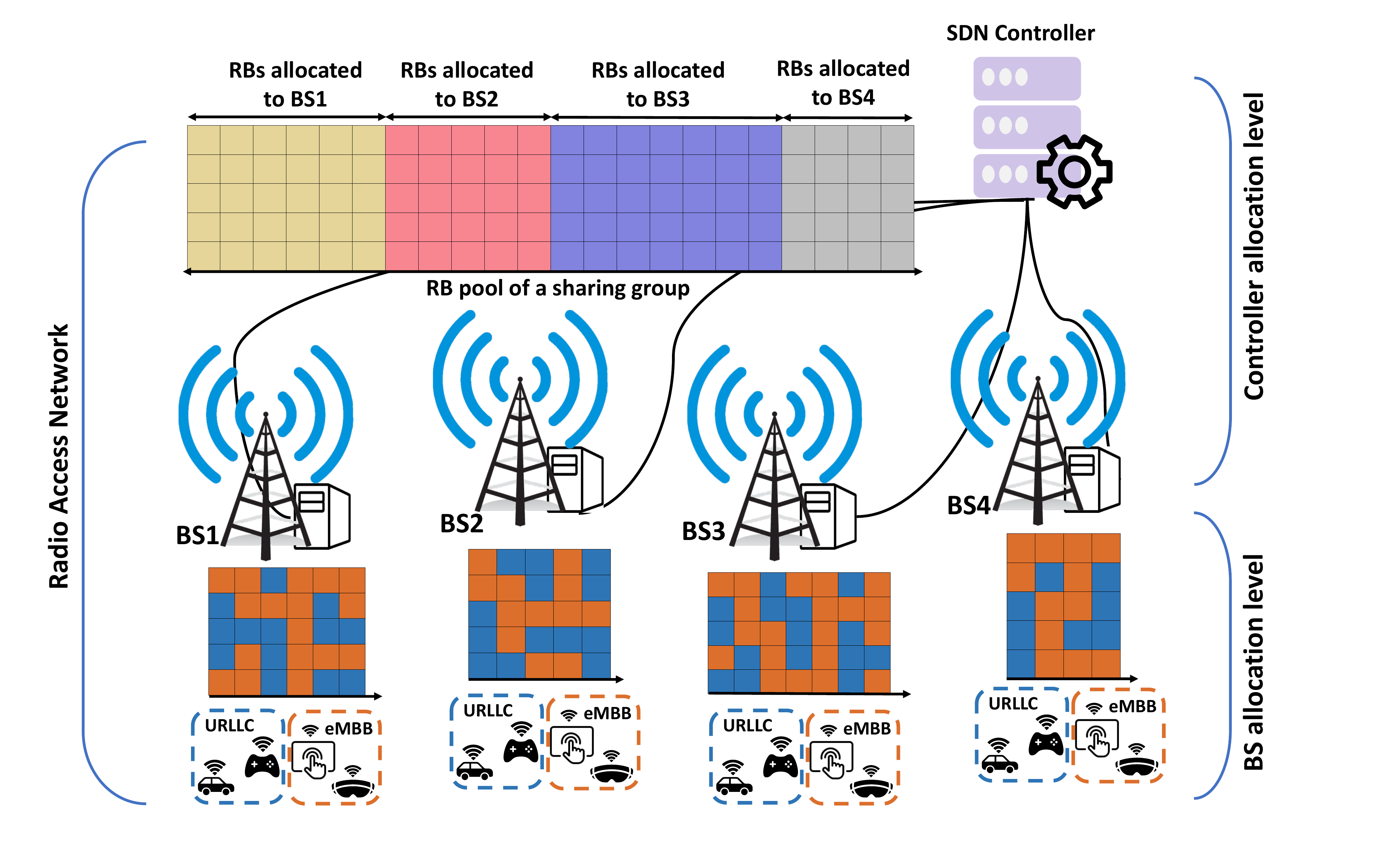}
	\caption{Resource block allocation procedure}
	\label{fig:1}
\end{figure}

Unlike the cloud RAN (C-RAN) architecture, which presents major challenges in its deployment with a multi-access edge computing (MEC) environment in terms of maintaining service availability and leveraging MEC resource ~\cite{etsi}, SDN can be recognized as an important 5G RAN enabler in a fog RAN (F-RAN) architecture. Specifically, co-deploying SDN with F-RAN increases the ability of exploiting radio resources using its global view of the network~\cite{SDN_FRAN}. The radio resource considered in this work is the resource block (RB), which is the minimum resource element that can be allocated to an end-user and it is identified by the frequency band and time slot pair. Each gNodeB should have enough RBs from the shared radio RBs pool to meet the requirements of its end-users. In a RAN slicing scenario, SDN can be used to allocate the appropriate RBs for each RAN slice in each gNodeB, depending on the radio resource availability and the services required by the end-users  of the corresponding gNodeB. A notable challenge here is how to achieve an optimal RBs allocation for each gNodeB to satisfy the quality of service (QoS) requirements of its end-users in different slices.

To solve this problem, a large variety of algorithms and schemes have been proposed in recent years \cite{rw1,rw2,rw10,rw3,rw4,rw5,rw7,rw8,rw9,ofdma} that mainly present either centralized resource allocation solutions or multi-level resource allocation solutions. However, we identify important gaps in these related-works. In the centralized solution approaches, the radio resource allocation decision relies only on a central entity (e.g., an SDN controller), which increases the signaling overhead in the network caused by frequent communications between the gNodeBs and the central entity, especially when the latter has to perform the resource allocation of the different gNodeBs. Therefore, the operation of allocating radio resources to end-users is expected to be performed in a large time-scale. On the other hand, in the multi-level resource allocation solution approaches, the SDN controller allocates, in the upper level, radio resources to RAN slices or gNodeBs. The latter are responsible, in the lower level and based on the pre-allocated resources by the SDN controller, for allocating radio resources to end-users. Although multi-level resource allocation solutions can efficiently allocate RBs to end-users, when the pre-allocated resources by the SDN controller are not sufficient to handle the required services, the low-level allocation operation can fail since it must wait for the next resource reservation update or immediately solicit the SDN controller for more resources, which can significantly reduce the QoS that the network operator is expected to provide.

In this paper, we fill these two gaps by proposing a two-level RB allocation mechanism. In the first level and in a large time-scale, the SDN controller allocates to each gNodeB a number of RBs from the common radio RBs pool, according to the gNodeB requirements. In the second level, each gNodeB schedules the pre-allocated RBs to its associated end-users in a short time-scale to satisfy their QoS requirements in terms of data rates and delay. This mechanism avoids frequent communications between gNodeBs and the SDN controller caused by abrupt and potentially unexpected fluctuations in wireless network traffic. To further reduce communications between gNodeBs and the SDN controller, a gNodeB can request additional RBs from other adjacent gNodeBs when its pre-allocated RBs are not sufficient, which allows to immediately respond to the requirements of the end-users. 

In this work, two types of slices are considered, each one is dedicated to a single 5G service, namely the eMBB service and the URLLC service. Then, we formulate the two-level RB allocation problem as an optimization problem where the objective is to maximize the total achievable data rate of eMBB and URLLC end-users. This objective has to be achieved subject to the ultra-low latency requirements of the URLLC services as well as to the minimum data rate requirements of the eMBB services. The proposed mechanism to solve this optimization problem will: (1) reduce the signaling overhead between the gNodeBs and the SDN controller, and (2) quickly provide the required RBs for different slices to meet the services requested by the corresponding end-users.

The novelty of this work lies in two main parts. In the first part, we define the two-level radio RB allocation problem using integer programming and we analyze its NP-hardness. In the second part, we propose a single-agent multi-agent reinforcement learning (SAMA-RL) framework to solve the two-level radio RB allocation problem. Therefore, the proposed RL-based RAN slicing framework is dynamic since RL algorithms can adapt their RB allocation policies permanently according to  different factors that mainly include the density of the end-users in the network, the requirements of the eMBB and the URLLC services, and the transmission conditions of the wireless channel. In addition, the proposed RAN slicing approach is performed in two-time scales to allocate resources for two levels including the SDN controller level and the gNodeBs level. Precisely, the proposed two-time scales RAN slicing approach is supervised by an SDN controller to manage the RB allocation policies dynamically, which means that the programmability provided by the SDN controller enables an automatic resource management. The main contributions of this paper are summarized as follows:
\begin{itemize}
    \item We use mathematical programming techniques to model the global RBs allocation for eMBB and URLLC end-users as a non-linear binary program and study its NP-hardness.
    \item Due to the NP-completeness result, obtaining an optimal solution to the RB allocation problem is computationally expensive. Thus, we model each level of the RB allocation problem as a Markov decision process (MDP).
    \item To fairly partition RBs between gNodeBs according to their requirements, we design a single agent RL-based algorithm to partition the RBs between gNodeBs in the first allocation level.
    \item In the gNodeBs level, based on the pre-allocated RBs by the SDN controller to gNodeBs, we propose a multi-agent deep Q-learning (DQL) approach to allocate RBs to eMBB and URLLC end-users and perform RBs sharing between gNodeBs.
    \item We evaluate the performance of our proposed mechanism against  benchmark algorithms and perform extensive simulations to show the superiority of the proposed framework.
\end{itemize}
The rest of the paper is organized as follows. Section II discusses analyzes the related works. Section III presents the system model. Section IV formulates the RB allocation problem as a mathematical program and studies its NP-hardness. Section V presents the proposed learning solutions. Section VI highlights the performances of the proposed mechanism and discusses the obtained results.  Finally, section VII concludes the paper.

\section{Related Work}
In \cite{rw1}, the authors propose a framework for RAN slicing by using two machine-learning approaches: (i) Long short-term memory (LSTM) is used to predict the resources that should be allocated to a slice in a large time-scale (ii) a multi-agent RL-algorithm, known as asynchronous actor-critic agent (A3C), is exploited for on-line resource scheduling of RAN slices. However, the proposed framework considers only eMBB end-users and the slicing for individual end-users is not considered. In \cite{rw2}, the controller reserves resources for URLLC and eMBB slices at each base station considering the minimum resource requirement of each slice. Then, to meet the required QoS and increase the resource utilization utility of slices, a deep RL (DRL) algorithm is executed for each slice on each base station to dynamically update the allocated RBs based on the reserved resources. Similarly, the authors of \cite{rw10} improve the user QoS satisfaction and resource utilization utility by adjusting the resource provided to individual RAN slices. They leverage DQL approach with the dueling DQN algorithm to solve the slice resource provisioning problem. However, if the resources reserved for a given slice are not sufficient to handle the required services, the slice must wait for the next resources reservation update since the slice cannot occupy more resources than those assigned by the controller. The authors of \cite{rw3} present a dynamic radio resource slicing scheme for a two-tier heterogeneous wireless network to determine the optimal bandwidth slicing ratios for slices. An alternative concave search algorithm is designed to solve the maximum network utility optimization problem. Although the proposed scheme satisfies the QoS requirement of machine-type and data slices, it cannot support URLLC slices which need much lower latency. Also, resources for each slice are expected to be updated only in a large time-scale. In \cite{rw4}, a two-level SDN-based radio resource allocation framework is designed to improve RAN slicing over different time scales. In a large time-scale, the SDN controller allocates RBs to gNodeBs, while in a small time-scale the pre-allocated RBs are scheduled by each gNodeB to eMBB and URLLC users. Moreover, each gNodeB can borrow RBs from other gNodeBs when the pre-allocated RBs are insufficient. However, the interactions between the base stations to share the RBs are not described. In meeting the eMBB, URLLC and mMTC slice user requirements, the authors of \cite{rw5} formulate the user-association and resource allocation problem as a maximum utility optimization problem. The optimization problem is decomposed in two sub-problems using the hierarchical decomposition method. The base station-slice user association sub-problem is solved by many-to-one matching game and a genetic algorithm is adopted to solve the dynamic resource allocation sub-problem. The authors of \cite{rw7} present a dynamic framework to allocate radio resources for two types of vehicular network services, i.e., delay-sensitive service and delay-tolerant service. The radio resource allocation problem is jointly formulated with that of computing resource allocation as an optimization problem. To solve this problem, they use a two-layer constrained RL algorithm. Based on the proposed RL algorithm, the SDN controller is responsible for allocating resources to slices at all the base stations. In order to provide their mobile users the access to the virtual computation and communication slices, multiple service providers compete in \cite{rw8} to orchestrate channel access opportunities. The authors model such resource allocation problem as a non-cooperative stochastic game and leverage a deep learning approach based on double deep Q-network to approximate the optimal allocation strategy. In \cite{rw9}, a two-step RAN slicing framework is proposed to improve bandwidth utilization. In the first step, the framework selects a set of users whose QoS can be satisfied simultaneously. In the second step, each admissible user is associated with a slice via a specific base station and a fraction of the base station bandwidth is allocated to it.

\section{System Model}
We consider an SDN-enabled 5G RAN architecture composed of a finite set of gNodeBs $\mathcal{B}= \{1,2,\ldots,B\}$. Two types of slices are considered, namely the URLLC slice and the eMBB slice, which are denoted by $s_u$ and $s_e$, respectively. Spectrum resources are represented as a shared RBs pool denoted by $\mathcal{K}=\{1,2,\ldots,K\}$, where each RB represents the minimum scheduling unit. The end-users $\mathcal{U}= \{1,2,\ldots,U\}$ are randomly located across the network area and where they are URLLC end-users and eMBB end-users. Each end-user $u \in \mathcal{U}$ is served by one gNodeB and belong to one slice, \ie $s_u$ or $s_e$. Each gNodeB $b \in \mathcal{B}$ has a set of associated end-users denoted by $\mathcal{U}_b$. We consider the orthogonal frequency division multiple access (OFDMA) downlink (DL) scenario, where the RBs are organized as a resource grid \cite{ofdma}. With OFDMA, the transmissions to end-users are scheduled in an orthogonal manner to reduce interference.

The RB allocation procedure is performed in two levels. In the first level, the SDN controller, since it has a global view of the network, allocates the RBs to the gNodeBs using the available RBs pool, in a large time-scale. We call this RB allocation level the SDN allocation level. The set of RBs assigned by the SDN controller to a gNodeB $b \in \mathcal{B}$ is donated by $\mathcal{K}_b \subseteq \mathcal{K}$. In the second level, the RB allocation procedure is performed by the gNodeBs to allocate the necessary RBs to each end-user. We call this RB allocation level the gNodeB allocation level. Each gNodeB $b\in\mathcal{B}$ allocates to each of its associated end-users, during a short time-scale, a number of RBs, from the pre-allocated RBs $\mathcal{K}_b \subseteq \mathcal{K}$, to satisfy the various QoS requirements in terms of data rate and latency of each end-user. To ensure the orthogonality of DL transmissions among end-users which are served by the same gNodeB $b\in \mathcal{B}$, each RB $k \in \mathcal{K}_b$ is exclusively assigned to one end-user $u_b \in \mathcal{U}_b$. Also, if $b\in \mathcal{B}$ borrows $k' \in \mathcal{K} \setminus \mathcal{K}_b$ to allocate it to one of its end-users, then this RB $k'$ should be unallocated. Thus, we define the assignment of an RB $k \in \mathcal{K}$ to an associated end-user $u_b \in \mathcal{U}_b$ with $b\in\mathcal{B}$ as a binary variable $x_{u_b}^{k}$, where: 
\begin{equation}\label{Xvar}
    x_{u_b}^{k}= 
    \begin{cases}
    1, & \text{if } k \in \mathcal{K} \text{ is assigned to } u_b \in \mathcal{U}_b, b \in \mathcal{B}, \\
    0, & \text{otherwise.}
    \end{cases}
\end{equation}
\begin{equation}\label{RBE}
    \sum_{u_b \in \mathcal{U}_b} x_{u_b}^k \leq 1, \forall k \in \mathcal{K}, \forall b\in \mathcal{B}.
\end{equation}
Eq.~\eqref{RBE} states that an RB must be allocated to only one end-user at a time, which meets the OFDMA constraints.

 In 5G NR, the scalable OFDM technology is a key innovation. The 3GPP 5G NR Release 15 specifications \cite{3gpp} state that the waveform is scalable in the sense that the subcarrier spacing of OFDM can be adapted to channel width. The choice of the spacing parameter depends on several factors, including the requirements of 5G services, \eg low latency for URLLC services and high data rate for eMBB services. Indeed, eMBB and URLLC services can be supported simultaneously on the same carrier by multiplexing two different numerologies, larger subcarrier spacing for URLLC services and lower subcarrier spacing for eMBB services. In our approach, the RB allocation operation is performed considering a given time-frequency resource grid model. In other words, for each numerology strategy, \ie a specific subcarrier spacing, such as 15 KHz, 30 KHz and 60 KHz, the appropriate trained model should be used to perform the RB allocation operation. Since the gNodeB is responsible for deciding which numerology strategy should be applied based on the channel state information, it can select the appropriate trained model for the selected numerology to allocate the RBs. Therefore, our approach can be applied with multiple 5G NR OFDM numerologies to allocate the RBs, provided that the appropriate trained model for each possible numerology strategy is available.

In this article, we assume that the gNodeBs have a (near) perfect knowledge of the channel state information \cite{CSI,CSI_1}.  The availability of (near) perfect channel state information at the gNodeB in a real 5G network can be justified when fading varies slowly over time and the mobility of the end-users is low since the wireless channel does not change rapidly, which is similar to our case. The channel state information can be accurately estimated in 5G networks using, for example, deep learning algorithms \cite{luo2018channel}.

The achievable data rate of the $u_b$-th end-user associated with the $b$-th gNodeB over the $k$-th RB and belonging to a slice $s \in \{s_u,s_e\}$  is defined as follows:

\begin{equation}\label{DR}
    r_{u_b,k}^s=W \log_2 \left(1+\frac{P_{u_b}^s G_{u_b,k}}{\sigma^2}\right),
\end{equation}
where $W$ denotes the bandwidth of an RB, $P_{u_b}^s$ is the transmission power of $b \in\mathcal{B}$ to end-user $u_b \in \mathcal{U}_b$ in slice $s\in\{s_u,s_e\}$, $G_{u_b,k}$ is the DL channel gain between $b\in\mathcal{B}$ and its associated end-user $u_b\in \mathcal{U}_b$, and $\sigma^2$ is the power of the additive white Gaussian noise (AWGN). We assume that the bandwidth and the downlink transmission power are the same for all RBs.

During the second level of resource allocation, if the SDN controller does not allocate sufficient RBs to a gNodeB, the latter can request additional RBs from other gNodeBs. In other words, we assume that a gNodeB can request additional RBs from other gNodeBs only if all its pre-allocated RBs are already assigned to its end-users. Mathematically, gNodeB $b$ can request additional RBs from other gNodeBs and allocate them to its associated end-users \textit{if and only if}
$\sum_{u_b\in\mathcal{U}_b}\sum_{k\in \mathcal{K}_b}x_{u_b}^{k} \geq|\mathcal{K}_b|$. This constraint can be defined as follows: 
\begin{equation}\label{Beq}
    x_{u_b}^{k'} |\mathcal{K}_b|\leq\sum_{u_b\in\mathcal{U}_b}\sum_{k\in \mathcal{K}_b}x_{u_b}^{k},   
\end{equation}
where $k' \in \mathcal{K}_{b'}$ is a borrowed RB by the gNodeB $b \in \mathcal{B}$ from gNodeB $b' \neq b \in \mathcal{B}$ and obviously $x_{u_b}^{k'} = 1$.

The total achievable data rate of end-user $u_b\in \mathcal{U}_b$ belonging to slice $s\in\{s_u,s_e\}$ and associated to gNodeB $b\in\mathcal{B}$ is defined as follows:
\begin{equation}\label{TDR}
    r_{u_b}^s=\sum_{k\in \mathcal{K}_b}x_{u_b}^{k} r_{u_b,k}^s+\sum_{\substack{b'\in \mathcal{B}\\
    b'\neq b}}\sum_{k'\in \mathcal{K}_{b'}} x_{u_b}^{k'} r_{u_b,k'}^s
\end{equation}

To calculate the delay for URLLC and eMBB traffic, we consider the following assumptions: (i) the arrival process of each gNodeB's packets follows a Poisson distribution, and (ii) the inter-arrival times of the packets are independent and follow an exponential distribution~\cite{f-sdn}. Accordingly, the queuing traffic model can be considered as an M/M/1 queuing system. In addition, the packet lengths for different slices are different but they are similar in a slice $s\in \{s_u,s_e\}$. By applying Little’s law, we calculate, in Eq.~\eqref{DL}, the average delay $d^s_{u_b}$ experienced by an end-user packet $u_b$ belonging to slice $s\in\ \{s_u,s_e\}$ and associated with $b \in \mathcal{B}$, where $\lambda_{u_b}^s$ is the packets arriving rate of $u_b \in \mathcal{U}_b$ and belonging to slice $s\in \{s_u,s_e\}$.
\begin{equation}\label{DL}
    d^s_{u_b}=\frac{1}{r_{u_b}^s-\lambda_{u_b}^s},
\end{equation}

 We choose the M/M/1 queuing system since it is widely used to characterize wireless communication systems, particularly in RAN slicing approaches \cite{rezazadeh2021zero,motalleb2019joint,rw7}. However, in some practical scenarios, the traffic becomes bursty and, therefore, the M/M/1 assumption (i.e., the packet arrival process of each gNodeB follows a Poisson distribution) may become too optimistic. In such scenarios, the gNodeB cannot be considered as an M/M/1 system and the queue cannot be modeled as a continuous-time Markov process. In this case, when the traffic is bursty, the queuing system can be modeled as a discrete-time Markov process \cite{xu2021performance,pappas2018performance}. As shown in \cite{shortle2018fundamentals}, many continuous-time Markov processes can be transformed into discrete-time Markov processes by observing only the state transitions. Therefore, since the M/M/1 queuing assumption can be seen as a continuous-time Markov process, the analyses and results obtained in this work are valid for most scenarios where the Poisson distribution cannot be applied. Note that, under different queuing assumptions, the mathematical analysis may be different and more complicated. Thus, for simplicity, we only assume the M/M/1 case.

\section{Problem Formulation and NP-Hardness}
\subsection{Problem Formulation}
The main question of RB allocation problem in RAN is how to derive a two-level optimal allocation of RBs to URLLC and eMBB end-users that meet their QoS requirements in terms of data rate and delay. For this purpose, the global RB allocation optimization problem is formulated as follows:
\begin{maxi!}[3]
	{x}{\sum_{b \in \mathcal{B}}\sum_{s\in \{s_u,s_e\}}\sum_{u_b \in \mathcal{U}_b} r_{u_b}^s \label{Obj}}
	{\label{opt}}{}
	\addConstraint{\sum_{k \in \mathcal{K}}x_{u_b}^{k}}{\leq K_{\text{max}}, \forall u_b \in \mathcal{U}_b, \forall b \in \mathcal{B} \label{C1}}
	\addConstraint{\sum_{u_b \in \mathcal{U}_b} x_{u_b}^k}{\leq 1, \forall k \in \mathcal{K}, \forall b  \in \mathcal{B}  \label{C2}}
	\addConstraint{x_{u_b}^{k'}\times |\mathcal{K}_b|}{\leq\sum_{u_b\in\mathcal{U}_b}\sum_{k \in \mathcal{K}_b }x_{u_b}^{k}, \forall k' \in \mathcal{K} \setminus \mathcal{K}_b, \forall u_b \in \mathcal{U}_b, \forall b\in \mathcal{B} \label{C3}}
	\addConstraint{r_{u_b}^{s_e}}{\geq \mathcal{R}_{min}, \forall u_b \in \mathcal{U}_b, \forall b \in \mathcal{B}  \label{C4}}
	\addConstraint {d_{u_b}^{s_u}}{\leq \mathcal{D}_{max}, \forall u_b \in \mathcal{U}_b, \forall b \in \mathcal{B} \label{C5}}
	\addConstraint{x_{u_b}^{k} \in \{0,1\}, \forall k \in \mathcal{K}, \forall u_b \in \mathcal{U}_b, \forall b \in \mathcal{B} \label{C6}}.
\end{maxi!}

The objective function in~\eqref{Obj} maximizes the total sum data rates of the URLLC and eMBB end-users. Constraints~\eqref{C1} guarantee a fair RBs allocation by forcing the number of RBs allocated to each end-user $u_b$ to not exceed a maximum number $K_{\text{max}}$. Constraints~\eqref{C2} respect the OFDMA constraints by ensuring that each RB is allocated to only one end-user at a time. Constraints~\eqref{C3} guarantee that a gNodeB $b$ can request additonal RBs from other gNodeBs if and only if all of its pre-allocated RBs are used by its end-users. Constraints~\eqref{C4} state that the data rate of the eMBB end-users must be greater than a minimum required threshold $\mathcal{R}_{min}$. Constraints~\eqref{C5} ensure that the delay of URLLC end-users cannot exceed a maximum required threshold $\mathcal{D}_{max}$. Finally, constraints~\eqref{C6}) list the optimization variables.

 Note that the dynamic resource allocation problem considers limited resources. In fact, the formulated RB allocation problem consists of maximizing the total achievable data rate of eMBB and URLLC end-users subject to QoS constraints and limited resources constraints represented by the set of constraints given in \eqref{C1}. These constraints guarantee that every end-user cannot be allocated more than a maximum number of resources. They limit the number of resources that each end-user can have and thus can allocate the remaining resources more efficiently and in a fair manner between end-users. The limitation of resources in the considered problem is also stated by the borrowing concept. Once a gNodeB is out of resources because all of its SDN-allocated resources are used, it can borrow other gNodeBs-resources.

Due mainly to the binary nature of the optimization variables and the non-linearity of the delay experienced by an end-user defined in Eq.~\eqref{DL}, the RB allocation problem is a non-linear binary programming problem. It is thus very challenging to solve~\eqref{opt} in general. In the sequel, we study its NP-hardness.

\subsection{NP-hardness}
Here, we denote the problem~\eqref{opt} by $P$. To prove that $P$ is NP-hard, we reduce the 0-1 knapsack problem \cite{Knapsack}, which is NP-hard, to an instance of $P$.

\begin{definition}[0-1 knapsack problem]
A 0-1 knapsack problem is defined as follows: given a set $\mathcal{N}$ of $n$ items, each one with its profit $p_i$ and weight $w_i$, and a knapsack of capacity $C$. Each item can be put into the knapsack or not (1 or 0). The objective of this problem is to find a subset $\mathcal{N}' \subseteq \mathcal{N}$ such that the total value of items $\sum_{n_i\in \mathcal{N}'} p_i$ is maximized and the total weight of the selected items is less than or equal to the knapsack capacity, i.e., $\sum_{n_i \in \mathcal{N}'} w_i \leq C$.
\end{definition}

\begin{theorem}
    $P$ is NP-hard.
\end{theorem}

\begin{IEEEproof}
    We prove the theorem by considering a restricted version of $P$. This shows that the problem is NP-hard in the general case as well. We consider the following restricted version of $P$:  
    \begin{itemize}
        \item there is only one gNodeB denoted by $b$ with its associated end-users set $\mathcal{U}_b$.
        \item there is only the eMBB slice.
        \item the RBs pre-allocated by the SDN controller $\mathcal{K}_b$ for $b$ are known, i.e., the first level of the RB allocation procedure is already performed by the SDN controller.
    \end{itemize}
    In this case, $P$ becomes equivalent to the following:
    \begin{maxi!}[3]
	    {x}{\sum_{u_b \in \mathcal{U}_b} r_{u_b}^{s_e} \label{Obj2}}
	    {\label{opt2}}{}
	    \addConstraint{\sum_{k \in \mathcal{K}_b}x_{u_b}^{k}}{\leq |\mathcal{K}_b|, \forall u_b \in \mathcal{U}_b \label{C2_1}}
	    \addConstraint{\sum_{u_b \in \mathcal{U}_b} x_{u_b}^k}{\leq 1, \forall k \in \mathcal{K}_b  \label{C2_2}}
	    \addConstraint{r_{u_b}^{s_e}}{\geq \mathcal{R}_{min}, \forall u_b \in \mathcal{U}_b, \label{C2_3}}
	    \addConstraint{x_{u_b}^{k} \in \{0,1\}, \forall k \in \mathcal{K}_b, \forall u_b \in \mathcal{U}_b, \label{C2_4}}.
    \end{maxi!}
    
    The mathematical statement of the problem in Eq.~\eqref{opt2} is equivalent to finding an allocation of the RBs set $\mathcal{K}_b$ to the eMBB end-user set $\mathcal{U}_b$ such that: (i) the sum data rates of the end-users is maximized, (ii) the maximum number of pre-allocated RBs to gNodeB $b$ does not exceed $K_{max}=|\mathcal{K}_b|$ \eqref{C2_1}, (iii) each RB is allocated to only one end-user at a time \eqref{C2_2}, and (iv) the data rate of the eMBB end-users must be greater than a minimum required threshold $\mathcal{R}_{min}$ \eqref{C2_3}.

    To reduce the 0-1 knapsack problem to~\eqref{opt2}, we let (i) the number of items $N$ be the number of eMBB end-users, (ii) the profit $p_i$ for item $i$ be the achievable data rate $r_{u_b}^{s_e}$ for eMBB end-user $u_b$, (iii) the weight $w_i$ for item $i$ be the allocated RBs for eMBB end-user $u_b$, and (iv) the knapsack capacity $C$ be the number of the pre-allocated RBs to gNodeB $b$. Problem $P$ is now clearly reduced to a knapsack problem. The knapsack capacity is respected and, thus, the number of allocated RBs to all eMBB end-users does not exceed $|\mathcal{K}_b|$.
    
    Since the reduction is clearly done in a polynomial-time and the 0-1 knapsack problem is NP-hard, we conclude that the restricted problem formulated in Eq.~\eqref{opt2} is also NP-hard, which proves the theorem.
\end{IEEEproof}

\section{Single-Agent Multi-Agent Reinforcement Learning Based RAN Resource Slicing}
\begin{figure}[t]
	\includegraphics[width=\linewidth]{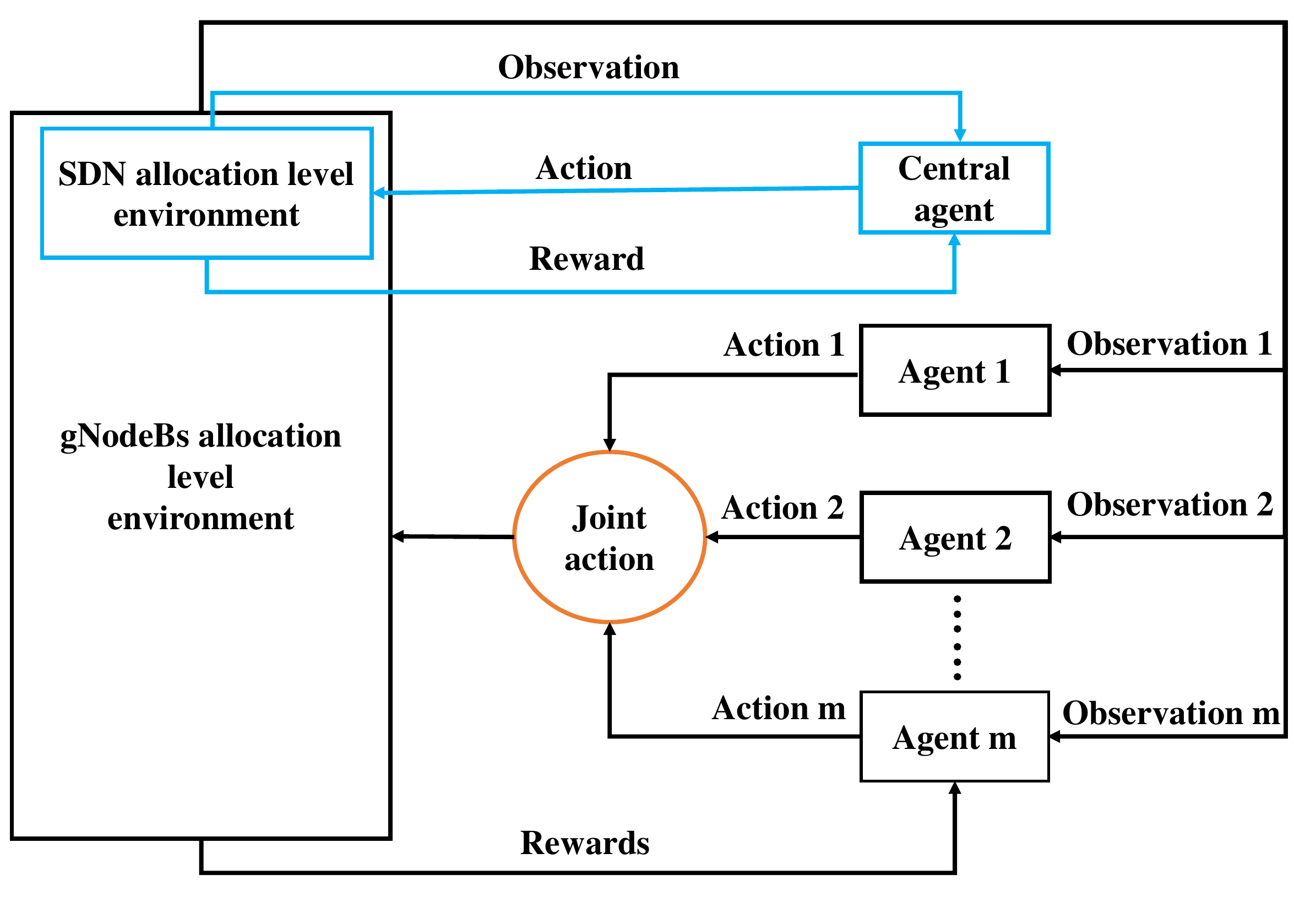}
	\caption{Single-agent multi-agent interaction with the MDP environment.}
	\label{fig:2}
\end{figure}

The resource allocation problem is very challenging to solve optimally in a two-level procedure, i.e., in the controller level and in the gNodeBs level. To overcome this challenge, we leverage machine learning techniques, particularly RL for performing RB allocation tasks due to its excellent capability to solve wireless network resource allocation problems in a computationally efficient manner~\cite{wang2020thirty}. We propose a single-agent multi-agent reinforcement learning (SAMA-RL), Fig.~\ref{fig:2}, framework to solve the two-level radio RB allocation problem.

More precisely, in the SDN allocation level, we adapt the well-known exponential-weight algorithm for exploration and exploitation (EXP3) \cite{MAB}. The SDN controller plays the role of an RL agent that performs the operation of allocating the RBs to the gNodeBs. In the gNodeB allocation level, a distributed multi-agent DRL approach is proposed. Each gNodeB, acting as a DRL agent, schedules the pre-allocated RBs by the SDN controller to its associated end-users and, if necessary, cooperate with the other gNodeBs to dynamically share the unexploited RBs between them. To do so, we apply a DQL \cite{DQL} model where each gNodeB acts as an independent agent. To avoid any confusion between the SDN controller agent and a gNodeB agent, the former will be called the central agent.

We opt for a classical RL algorithm, i.e., EXP3, in the SDN allocation level because this resource allocation level does not suffer from the curse of dimensionality problem. In fact, the state and action spaces are not large, which allows the central agent to learn in a reasonable time. EXP3 is a promising algorithm because it does not depend on any assumption related to the dynamicity of the SDN allocation system, which makes it relevant when the channel gains offered by RBs vary randomly. Moreover, EXP3 provides a good tradeoff between exploitation, the desire to make the best decision given current information, and exploration, the desire to try a new decision which may lead to better results. On the other hand, the gNodeB allocation level suffers from the curse of dimensionality due to the multi-agent scenario and presence of a large number of end-users, which leads the size of state space to grow significantly. Therefore, using a simple RL framework becomes computationally intractable.  To overcome this challenge, we resort to a distributed multi-agent DRL approach in the gNodeB allocation level. Indeed, the agents are capable of discovering meaningful information through an appropriate deep neural network architecture. Therefore, the agents can learn close-optimal policies.  In addition, DRL has been proven to be efficient in solving many resource allocation problems in wireless networks, such as energy scheduling \cite{zhang2017energy} and orchestration of edge computing and caching resources \cite{dai2019artificial}.

Before explaining in details the proposed SAMA-RL algorithms, we first model, in what follows, each RB allocation problem of each level as a Markov decision process (MDP) and we define the main components of each MDP in details.

\subsection{MDP formulation of the SDN allocation level}
The MDP of the first resource allocation level is given by the triplet $(\mathscr{S}_c, \mathscr{A}_c, \mathscr{R}_c)$ where $\mathscr{S}_c$ represents the state space, $\mathscr{A}_c$ designates the action space and $\mathscr{R}_c$ is the reward function.
\subsubsection{The state space} ~

As discussed previously, the central agent is incorporated into the SDN controller which has a global view of the network. In fact, the central agent's state contains information about its previously allocated RBs to the gNodeBs and other global parameters of the network. More precisely, the state space of the central agent is composed of the triple $\mathscr{S}_c$ given as follows:
\begin{equation}\label{ctrlState}
    \mathscr{S}_c = (\mathcal{B},\mathcal{K},\boldsymbol{\mathcal{K}_h}),
\end{equation}
where $ \boldsymbol{\mathcal{K}_h}= \{ \mathcal{K}_b, \forall b \in \mathcal{B} \}$ represents the allocated RBs previously to each gNodeB. Also, the central agent' state includes the sets $\mathcal{B}$ and $\mathcal{K}$ that represent the sets of gNodeBs and RBs, respectively.  

\subsubsection{The action space} ~

Since the central agent performs the RB allocation in a large time-scale, it has to decide which subset $\mathcal{K}_b \subseteq \mathcal{K}$ of RBs it should allocate to each gNodeB $b \in \mathcal{B}$. Note that the central agent has to make sure that an RB is assigned to only one gNodeB each time it performs the RB allocation. Therefore, the action space of the central agent, denoted by $\mathscr{A}_c$, is defined as follows:

\begin{equation}\label{ctrlAction}
    \mathscr{A}_c = \{0,1\}^{|\mathcal{K}| \times |\mathcal{B}|},
\end{equation}
where an action $\boldsymbol{a_c} \in A_c$ is given by the row vector $ [a_1^1,\ldots,a_1^{|\mathcal{K}|},a_2^1,\ldots,a_2^{|\mathcal{K}|},\ldots,a_{|\mathcal{B}|}^1,\ldots,a_{|\mathcal{B}|}^{|\mathcal{K}|}] $. If an element $a_b^k$ of vector $\boldsymbol{a_c}$ is equal to 1, it means that the central agent has decided to allocate RB $k$ to gNodeB $b$. Also, to avoid assigning the same RB $k$ to several gNodeBs, the central agent has to make sure that the constraints $a_b^k \neq a_{b'}^k, \forall k \in \mathcal{K}$ and $b \neq b' \in \mathcal{B}$ are satisfied. 

\subsubsection{The reward function} ~

Once the central agent observes the environment through its current state, it chooses an action $\boldsymbol{a_c}$ from the set $\mathscr{A}_c$. After choosing an action from the action space $\mathscr{A}_c$, the central agent receives a reward $\mathscr{R}_c$. Since the objective is to maximize the total sum-rate, the objective of the central agent has to be related to the sum-rate of the network given in Eq.~\eqref{Obj}. We define the reward as how the assigned RBs to gNodeBs affect the achieved data rate of the entire system. In other words, the more the total data rate of the system is higher, the better is the action chosen by the central agent.
Thus, the reward of the central agent is given by Eq.~\eqref{Obj}, where:

\begin{equation}\label{ctrlReward}
    \mathscr{R}_c = \sum_{b \in \mathcal{B}}\sum_{s\in \{s_u,s_e\}}\sum_{u_b \in \mathcal{U}_b} r_{u_b}^s    
\end{equation}

According to Eq. \eqref{ctrlReward}, the value of $\mathscr{R}_c$ depends on the RB allocation decisions in the gNodeB level. Indeed, the central agent should wait for the results of the second resource allocation level in order to efficiently explore its environment. Therefore, the RB allocation is performed in a joint manner between the two allocation levels.

\subsection{MDP formulation of the gNodeB allocation level}
The MDP of the second resource allocation level is given by the triplet $(\mathscr{S}_b, \mathscr{A}_b, \mathscr{R}_b)$ where $\mathscr{S}_b$ represents the state space of agent $b$, $\mathscr{A}_b$ designates the action space of agent $b$ and $\mathscr{R}_b$ is the reward function of agent $b$.
\subsubsection{The state space} ~

As illustrated in Fig. 2, we propose a multi-agent DQL algorithm, with each gNodeB acts as an agent, to allocate the required RBs to eMBB and URLLC end-users and performs RBs sharing between gNodeBs, if necessary. Indeed, based on the chosen action by the central agent in the SDN allocation level, each agent observes its local state $\mathscr{S}_b$. In our model, we consider that the controller communicates to each agent the information about the RBs allocated to all other agents. Accordingly, each agent $b$'s state $\mathscr{S}_b, \forall b \in \mathcal{B}$, is given by the following tuple:
\begin{equation}\label{BSState}
    \mathscr{S}_b = (\mathcal{G}_b,\mathcal{U}_b,\mathcal{K}_b,\mathcal{R}_{min},\mathcal{D}_{max}),
\end{equation}
where $\mathcal{U}_b$ and $\mathcal{K}_b$ represents the set of associated end-users with agent $b$ and the set of pre-allocated RBs to agent $b$ respectively. The term $\mathcal{G}_b = (G_{{u_b},k}:u \in U_b, k \in \mathcal{K})$ represents the DL channel gain between agent $b$ and its associated end-users in each RB $k \in \mathcal{K}$. The channel gains $\mathcal{G}_b$ can be easily collected by agent $b$ as follows: (i) agent $b$ broadcasts pilot signals to all of its associated end-users. Then, each end-user estimates the channel state information and sends it back to the corresponding agent through a feedback channel. The estimation of $\mathcal{G}_b$ between each associated end-user over each RB in the system helps agent $b$ to take good decisions, especially to borrow the needed RBs from the other agents. The state $\mathscr{S}_b$ also includes the minimum data rate threshold $\mathcal{R}_{min}$ and the maximum delay threshold $\mathcal{D}_{max}$ that corresponds to the requirements of eMBB and URLLC slices, respectively.

\subsubsection{The action space} ~

In the gNodeB allocation level, each agent takes actions according to its own allocation policy. In fact, the agent has to (i) assign the pre-allocated RBs to its associated end-users and (ii) request additional RBs from other agents when its pre-allocated RBs are not sufficient. Also, it is assumed that each agent is aware of the number of RBs present in the entire system. This information is acquired as follows: in each round of the SDN allocation level, the controller communicates with each agent the information on the RBs allocated to the other agents. Therefore, we define the space action of agent $b$, denoted by $\mathscr{A}_b$, as follows: 
\begin{equation}\label{BSAction}
    \mathscr{A}_b = \{0,1\}^{|\mathcal{K}| \times |\mathcal{U}_b|},
\end{equation}
where an action $\boldsymbol{a_b} \in \mathscr{A}_b$ is given by the row vector $ [a_1^1,\ldots,a_1^{|\mathcal{K}|},a_2^1,\ldots,a_2^{|\mathcal{K}|},\ldots,a_{|\mathcal{U}_b|}^1,\ldots,a_{|\mathcal{U}_b|}^{|\mathcal{K}|}] $. Note that a vector $\boldsymbol{a_b}$ is equivalent to an association matrix $[x_{u_b}^k]$ because each element $a_{u_b}^k$ of vector $\boldsymbol{a_b}$ corresponds to the assignment of an RB $k \in \mathcal{K}$  to an associated end-user $u_b \in \mathcal{U}_b$. In other words, if an element $a_{u_b}^k$ of vector $\boldsymbol{a_b}$ is equal to 1, it means that agent $b$ has decided to allocate RB $k$ to end-user $u_b$. Also, if all elements $ a_{u_b}^{k'}, \forall k' \in \mathcal{K} \setminus \mathcal{K}_b$, of vector $\boldsymbol{a_b}$ are equal to 0, it means that agent $b$ does not request additional resources from the other gNodeBs. When constructing the action space $\mathscr{A}_b$ of agent $b$, the constraints \eqref{C1}, \eqref{C2} and \eqref{C3} should be respected. By applying these constraints, we significantly reduce the action space $\mathscr{A}_b$. As a result, the exploration phase is significantly improved to discover better strategies, which considerably accelerates the learning process of agent $b$. Further, considering that agents should cooperate to dynamically share the unexploited RBs among themselves, each agent communicates its chosen action to the others. Therefore, each agent $b$ forms a joint action $\boldsymbol{a}=(\boldsymbol{a}_b,\boldsymbol{a}_{-b})$ where $\boldsymbol{a}_{-b}$ denotes the actions chosen by the other agents. 

\subsubsection{The reward function} ~

Multi-agent reinforcement learning methods seek to learn a policy that achieves the maximum expected total reward for all agents. Indeed, the learning process of all agents is driven by the reward function. In our model, the main objective is to maximize the total achievable data rate of the entire system to meet the QoS requirement of eMBB and URLLC end-users. Therefore, the reward function of agent $b$ relates to its total sum-rate subject to the ultra-low latency requirements of the URLLC services as well as to the minimum data rate requirements of the eMBB services.

The reward of agent $b$ depends on whether or not it has successfully allocated the needed RBs to its associated end-users. An RB allocation operation is considered to be feasible if the chosen action $\boldsymbol{a_b}$ satisfies the constraints \eqref{C1}, \eqref{C2}, \eqref{C3}, \eqref{C4} and \eqref{C5}, otherwise it is considered as an infeasible operation. The constraints (7b), (7c) and (7d) are already verified in the action space construction phase. However, as an allocation operation can require borrowing some RBs from other agents, it is necessary to verify if a borrowed RB is: (i) exploited by its owner and (ii) also chosen by at least one agent other than its owner. Since each agent forms the joint action $\boldsymbol{a}$, (i) and (ii) can be easily verified by agent $b$. If at least one of them is correct, the constraint \eqref{C3} is not satisfied and thus the action chosen by agent $b$ is not feasible. As a result, the individual reward of agent $b$, denoted by $\mathscr{R}_b$ is expressed as follows:
\begin{equation}\label{BSReward}
    \mathscr{R}_b= 
    \begin{cases}
    \sum\limits_{s\in \{s_u,s_e\}}\sum\limits_{u_b \in \mathcal{U}_b} r_{u_b}^s, & \text{if } \boldsymbol{a_b} \text{ is feasible}, \\
    -1, & \text{if } \boldsymbol{a_b} \text{ is not feasible}.
    \end{cases}
\end{equation}

When an action $\boldsymbol{a_b} \in \mathscr{A}_b$ is not feasible, it is penalized with a negative reward, $\mathscr{R}_b=-1$, to prevent the agent from choosing infeasible actions in the future. 

\subsection{Single-agent EXP3 algorithm}
In order to solve the SDN allocation level problem, we adopt an online learning algorithm that is based on the multi-armed bandit (MAB) approach \cite{MAB}. In MAB, a player needs to choose, at each round of the game, one arm from a finite set of arms, each characterized by an unknown reward, with the objective of maximizing his expected cumulative reward. The single-agent MDP is modeled as a MAB as follows. The central agent, i.e., the SDN controller, represents the player and the set of arms is given by its action space $\mathscr{A}_c$. We propose the EXP3 algorithm as a popular bandit strategy to solve the SDN allocation level problem \cite{z_iot}.  The SDN controller runs the EXP3 algorithm where each action is assigned a weight to evaluate how good the action is for the SDN controller, \ie the higher the weight of an action, the better the action is. At the beginning of the algorithm, the weights of all actions are uniformly distributed. Then, the algorithm iterates several rounds. For each round, the SDN controller:

\begin{enumerate}
    \item calculates the probability of choosing each action, which is proportional to its weights;
    \item chooses an action according to the probability distribution calculated previously and receives a reward; and 
    \item uses the received reward to update the weights of each action by applying an exponential weighting scheme. The advantage of such a scheme is that it rapidly increases the probability of good actions, while rapidly reducing the probability of bad actions.
\end{enumerate}
The pseudo-code of the EXP3 algorithm is presented in Algorithm~\ref{alg:Exp3}.


In detail, the EXP3 algorithm takes as input an exploration parameter $\alpha \in [0,1]$ that controls the desire to choose an action uniformly at random. It starts by assigning a weight $\psi_i$ to each action $\boldsymbol{a_{c,i}}$, which is initialized to 1. Then, it iterates the rounds. For each round, the central agent calculates the probability $\pi_i$, given in Eq.~\eqref{PiExp3}, of choosing action $\boldsymbol{a_{c,i}}$. Based on $\pi$, it selects an action $\boldsymbol{a_{c,i}}$ and receives a reward $\mathscr{R}_{c,i}$. The obtained reward is scaled to the range $[0,1]$ and it is denoted by $\mathscr{\bar{R}}_{c,i}$. Since $\mathscr{R}_{c,i}$ is the total achievable data rate of the system, it can be scaled using Eq.~\eqref{ScaledReward}. After scaling the obtained reward, the central agent calculates an estimated reward $\mathscr{\hat{R}}_{c,i}$ using Eq.~\eqref{EstimatedReward}. The idea behind estimating the reward in such manner is to compensate for a potentially low probability of obtaining the observed reward. Finally, the weights are updated as $\psi_i = \psi_i \exp{(\alpha \mathscr{\hat{R}}_{c,j}/|\mathscr{A}_c|)}$.
\begin{equation}\label{PiExp3}
    \pi_i = (1-\alpha)\frac{\psi_i}{\sum_{j=1}^{|\mathscr{A}_c|}\psi_j} + \frac{\alpha}{|\mathscr{A}_c|}
\end{equation}
\begin{equation}\label{ScaledReward}
    \mathscr{\bar{R}}_{c,i} = 1 - \frac{1}{(1+\mathscr{R}_{c,i})}
\end{equation}
\begin{equation}\label{EstimatedReward}
    \mathscr{\hat{R}}_{c,i} =  \frac{\mathscr{\bar{R}}_{c,i}}{\pi_i}
\end{equation}

\begin{algorithm}[t]
	\caption{EXP3-based SDN allocation level}
	\label{alg:Exp3}
	\begin{flushleft}
	    Parameters: $\alpha \in [0,1]$ \\
		Initialize $\psi_i = 1$ for all $i \in \{1,2,\ldots,|\mathscr{A}_c| \}$
	\end{flushleft}
	\begin{algorithmic}[1]
		\FOR{\textit{each round}}
		    \FOR{$i=1,2,\ldots,|\mathscr{A}_c|$}
		       \STATE Calculate $\pi_i$ using Eq.~\eqref{PiExp3}
		    \ENDFOR
		    
    		\STATE Select an action $\boldsymbol{a_{c,i}}$ according to
    		$\pi_i$
    		\STATE Receive reward $\mathscr{R}_{c,i}$
    		\STATE Calculate $\mathscr{\bar{R}}_{c,i}$ using Eq.~\eqref{ScaledReward}
    		\FOR{$j=1,2,\ldots,|\mathscr{A}_c|$}
    		    \STATE $\mathscr{\hat{R}}_{c,j} \leftarrow \mathscr{\Bar{R}}_{c,j}/\pi_i \cdot \mathbbm{1}_{j=i}$ 
    		    
    		    \STATE $\psi_i \leftarrow \psi_i \exp{(\alpha \frac{\mathscr{\hat{R}}_{c,j}}{|\mathscr{A}_c|})}$
    		\ENDFOR
		\ENDFOR
	\end{algorithmic}	
\end{algorithm}

The EXP3 algorithm is simple to implement and does not require enormous computational complexity since it simply updates the weights of choosing actions by increasing or decreasing their probabilities according to their performance.
 Note that the EXP3 algorithm is an online learning algorithm that enables the SDN controller to increase or decrease the weight of the actions according to the feedback received from the gNodeBs. Precisely, the SDN controller selects an action $\boldsymbol{a_{c,i}}$ to allocate the RBs to the gNodeBs. Then, it waits for the results of the second RB allocation level, which is performed by the gNodeBs. Once the gNodeBs have assigned the RBs, chosen by the SDN controller, to their end-users, each gNodeB: 1) calculates its achieved data rate, and 2) communicates this information to the SDN controller. The total sum-data rate of all gNodeBs will be the reward received by the SDN controller after choosing action the $\boldsymbol{a_{c,i}}$, that will be used by the EXP3 algorithm to update the actions' weights.

\subsection{Multi-agent deep Q-Learning algorithm}
 The DQL algorithm \cite{DQL} extends the classical Q-learning RL \cite{QL} algorithm by approximating the Q-function using a deep neural network known as deep Q-network (DQN). In order to be used as an efficient non-linear approximator, such a network must be trained to observe the state of the agent and learn weights in order to play actions that yield the highest rewards. Once the DQN is properly trained, it is exploited by the agent to take actions based on the observed state~\cite{z_dqn}. To solve the multi-agent MDP model, we propose a multi-agent DQL algorithm. This algorithm consists of two main phases: the training phase and the implementation phase. In the training phase, each agent trains a deep neural network (DNN) in an offline manner using a large amount of experiences (collected dataset). In the implementation phase, each agent chooses actions in an online manner using its trained model. We describe, in the following, the training and implementation phases of the proposed multi-agent DQL approach.
\subsubsection{The training phase of DQL}~

DQN approximates the Q-value function $\mathscr{Q}(s,a)$ through a neural network that performs a mapping between states and actions. In other words, this network returns, for any given state-action pair, the estimated Q-value $\mathscr{Q}(s,a;w)$, where $w$ represents the parameters of the network (i.e., the weights). In order to improve the learning performance, DQN introduces the experience replay memory strategy that overcomes the learning stability issues. Indeed, this strategy stores the agent’s experiences that include state transitions, actions and rewards, and then randomly samples from these experiences to perform Q-learning. As a result, the experience replay memory strategy reduces the correlation between the training samples, which prevents the optimal policy from being conducted to a local minimum. Although DQN can be effective, it still suffers from the problem of overestimating action Q-values. Double DQN (DDQN) \cite{DDQN} is proposed to mitigate this limitation and improve learning performance. The idea behind DDQN is to decouple action selection from evaluation. To achieve this, two neural networks are used, a main Q-network that selects an action and a target Q-network that calculates the Q-value of the selected action. In our DQL-based MARL algorithm, each agent $b$ has a DDQN that takes the current state as input and outputs the Q-value function of all actions. The training phase of the proposed multi-agent DDQN is given in Algorithm~\ref{alg:trainingPhase}. 
\begin{algorithm}[t]
	\caption{DQL Algorithm Training Phase}
	\label{alg:trainingPhase}
	\begin{flushleft}
		\textbf{Input:} Agents and environment\\
		\textbf{Output:} Trained DDQNs
	\end{flushleft}
	Start simulator: generate end-users and network parameters; \\
	Initialize for each agent $b$: the main DQN, the target DQN and the replay memory $\mathscr{M}_b$;
	\begin{algorithmic}[1]
        \FOR{\textit{each episode}}
            \STATE Reset and build the agents' environment;
            \FOR {\textit{each step}}
                \FOR {\textit{each agent} $b$}
                    \STATE Get observation $\mathscr{S}_b$;
                    \STATE Choose an action $\boldsymbol{a}_b$ using $\epsilon$-greedy;
                \ENDFOR
                \FOR {\textit{each agent} $b$}
                    \STATE Obtain the joint action $\boldsymbol{a}$ and receive reward $\mathscr{R}_b$;
                    \STATE Obtain the next observation $\mathscr{S}_b'$;
                    \STATE Store the experience $exp_b$\ in replay buffer $\mathscr{M}_b$;
                    \IF{\textit{batch size}}
                        \STATE Randomly sample a mini-batch from $\mathscr{M}_b$;
                        \STATE Calculate target Q-value;
                        \STATE Calculate loss between the main network and the target network;
                        \STATE Update the parameters of the main network using gradient descent to minimze loss;
                    \ENDIF
                    \IF{\textit{target step}}
                        \STATE Update the target network parameters;
                    \ENDIF
                \ENDFOR
            \ENDFOR
        \ENDFOR
    \end{algorithmic}
\end{algorithm}

In detail, the training phase requires as input the environment of each agent which includes gNodeBs, end-users, pre-allocated RBs, service requirements and channel state information. As output, it returns the trained DQN of each agent. The training starts by : (1) generating the network parameters, the end-users including their positions on the grid, the service required (i.e., eMBB or URLLC) and their packet sizes, and (2) initializing the DQN of each agent. Next, DQL iterates the episodes. At the beginning of each episode, each agent’s environment is built by updating the end-user locations and the channel coefficients (i.e., large scale fading). In each step, each agent $b$ observes the current state $\mathscr{S}_b$ of its environment and takes an action $\boldsymbol{a_b}$ from its action space $\mathscr{A}_b$ by using the $\epsilon$-greedy policy. With the $\epsilon$-greedy policy, an agent selects an action randomly or using the Q-network. Precisely, this policy chooses the action with the highest Q-value with probability $1 - \epsilon$, where the exploration rate $\epsilon$ represents the probability that an agent will explore its environment rather than exploit it. As we go forward (i.e., after each step), each agent learns more about its environment and $\epsilon$ decays by some rate, so that exploring the environment becomes less probable. Once all agents choose their action following the $\epsilon$-greedy policy, they communicate them to each other. Accordingly, each agent $b$ forms its joint action, calculates its reward $\mathscr{R}_b$ using Eq. \eqref{BSReward} and moves to a new state $\mathscr{S}_b'$. Next, the obtained tuple $(\mathscr{S}_b,\boldsymbol{a_b},\mathscr{R}_b,\mathscr{S}_b')$, called agent’s $b$ experience and denoted by $exp_b$, is stored in its replay memory $\mathscr{M}_b$. In practice, since the size of the replay memory is limited to a defined threshold $\mathcal{M}$, only the last $\mathcal{M}$ experiences can be stored. After storing enough experiences, each agent samples a random mini-batch from its replay memory. Note that the size of the replay memory should be large enough to reduce the correlation between the data that will be sampled from it. The obtained dataset is used by the agent to perform the training. With the objective of minimizing the loss function, given by Eq. \eqref{loss}, the main Q-network is used to approximate the Q-value function while the target Q-network is used to outputs the target Q-value. 

\begin{equation}\label{loss}
    L_b(w_b)= \mathbbm{E}[(y_b - \mathscr{Q}(\mathscr{S}_b,\boldsymbol{a_b};w_b))^2],
\end{equation}
where $\mathscr{Q}_b(\mathscr{S}_b,\boldsymbol{a_b};w_b)$ is the approximated Q-value function given by the main Q-network of agent $b$ with weight parameter $w_b$ and $y_b$ denotes the target Q-value and it is given as follows:
\begin{equation}\label{targetV}
    y_b= \mathscr{R}_b + \gamma\mathscr{Q}(\mathscr{S}_b,\underset{\boldsymbol{a_b}}{max} \,\{ \mathscr{Q}(\mathscr{S}_b,\boldsymbol{a_b};w_b)\};w^{-}_b),
\end{equation}
where $0 \leq \gamma \leq 1 $ is called the discount factor, $w^{-}_b$ is the weight parameter of the target Q-network. Note that the value of $y_b$ is not necessarily the largest Q-value in the target Q-network, which allows to avoid chossing an overestimated action. 

After calculating the loss function, each agent performs a gradient descent to update the parameters of the main Q-network. Finally, the parameters of the target Q-network are updated, at each fixed target step, by copying the parameters of the main Q-network.

Since the learning of the DDQNs is computationally heavy, the training phase of the DQL algorithm is performed in an offline manner. Accordingly, the training can be conducted using a large amount of dataset resulting from different network topologies and channel conditions. 
\begin{algorithm}[t]
	\caption{DQL Algorithm Implementation Phase}
	\label{alg:implementationPhase}
	\begin{flushleft}
		\textbf{Input:} The trained DDQNs\\
		\textbf{Output:} RB allocation for end-users
	\end{flushleft}
	Load the DDQN of each agent;
	\begin{algorithmic}[1]
        \FOR{\textit{each episode}}
            \STATE Reset and build the agents' environment;
            \FOR {\textit{each step}}
                \FOR {\textit{each agent} $b$}
                    \STATE Obtain observation $\mathscr{S}_b$;
                    \STATE Choose $\boldsymbol{a}_b$ that maximize the Q-function;
                \ENDFOR
                \STATE Obtain the joint action $\boldsymbol{a}$;
                \STATE Find a solution to RB allocation for end-users;
            \ENDFOR
        \ENDFOR
    \end{algorithmic}        
\end{algorithm}
\subsubsection{The implementation phase of DQL}~

After the training phase, the parameters of the main Q-networks are used to find an RB allocation solution for the end-users in the online implementation phase of the DQL algorithm. The implementation phase is presented in Algorithm~\ref{alg:implementationPhase}. This phase uses the trained DDQNs of the agents. At the beginning of each episode, it builds the environment of each agent. Then, for each step, when a new state is observed, each agent $b$ selects the action that maximizes the Q-value of current state. Once all agents have chosen their actions, each agent can form a joint action. Accordingly, an RBs allocation solution is obtained.

\section{Simulation Results}
This section investigates the performance of the proposed two-level RB allocation mechanism through several simulated scenarios. 

\subsection{Experiment scenarios and setup}
We consider an SDN-enabled RAN architecture where the gNodeBs are deployed in a square of area $1 \text{Km}^2$. The end-users are uniformly distributed across the entire coverage area where each one of them is associated to only one gNodeB. Each end-user is assumed to be either an eMBB end-user or a URLLC end-user. For the sake of simplicity, the transmission power $P_{u_b}^s$ is the same for all $ u_b \in \mathcal{U}_b$ and $ b \in \mathcal{B}$. The key parameters of the simulations are summarized in Table \ref{tab:1}. In order to select an effective Q-network model, the training phase of the DQL algorithm is performed on a laptop with an Intel Core i7-8750H processor, 16 GB of RAM and NVIDIA GeForce GTX 1070 graphic card. We create and train the DDQNs of the agents using the PyTorch framework.   To select the best hyperparameters values for training the DDQN models, extensive simulations are performed. Indeed, we tested random combinations of hyperparameters in a fine-grained set of values chosen based on common settings in the literature \cite{rw8,9177109}. In particular, each DDQN consists of two fully connected hidden layers, each with 256 neurons. Rectified linear unit (ReLU) is used as the activation function to avoid the vanishing gradient problem in backpropagation, which accelerates the learning process. The Adam optimizer is used with a learning rate of 0.001 since it is computationally efficient. During the training process, the weights of the main Q-network are copied to the weights of the target Q-network every 1000 steps to avoid overestimating the Q-values. In addition, we consider end-users with low mobility, so the channel gains between a gNodeB and an end-user remain unchanged for a certain period of time. To do so, we fix the location of the end-user for a few training episodes, which helps the learning algorithm to better acquire the dynamics of the end-users and, at the same time, stabilize the training. The other hyperparameters of DDQN are given in Table \ref{tab:2}.

\begin{table}[t]
	\centering
	\caption{Simulation parameters.}
	\label{tab:1}
	\begin{tabular}{|C{6cm}|C{2cm}|}
		\hline
		
		\textbf{Parameter} & \textbf{Value} \\ \hline
		
		\multicolumn{1}{|l|} {Number of gNodeBs} & 2 \\\hline
		\multicolumn{1}{|l|} {Total number of end-users} & 8 \\\hline
		\multicolumn{1}{|l|} {Bandwidth of an RB} & 180 KHz \\\hline
		\multicolumn{1}{|l|} {Transmit power of gNodeB, $P_{u_b}^s$} & 30 dBm \\\hline
		\multicolumn{1}{|l|} {Power of AWGN, $\sigma^2$} & -114 dBm\\\hline
		\multicolumn{1}{|l|} {Packet arriving rate per end-user, $\lambda_{u_b}^s$} & 100 pachets/s\\\hline
		\multicolumn{1}{|l|} {Packet length for an eMBB \& URLLC end-user} & 400 \& 120 bits\\\hline 
		\multicolumn{1}{|l|} {Minimum data rate for eMBB end-user, $\mathcal{R}_{min}$} & 100 kbps\\\hline
		\multicolumn{1}{|l|} {Maxim delay for URLLC end-user, $\mathcal{D}_{max}$} & 10 ms\\\hline
        
	\end{tabular}
\end{table}
\begin{table}[t]
	\centering
	\caption{Retained hyper-parameters for DDQN.}
	\label{tab:2}
	\begin{tabular}{|C{5cm}|C{3cm}|}
		\hline
		
		\textbf{Hyper-parameter} & \textbf{Value}    \\\hline
		
		\multicolumn{1}{|l|} {Learning rate} & 0.001                       \\\hline
		\multicolumn{1}{|l|} {Epsilon/$\epsilon$-greedy} & 1               \\\hline
		\multicolumn{1}{|l|} {Discount factor} & 0.996                     \\\hline
		\multicolumn{1}{|l|} {$\epsilon$-min} & 0.01                       \\\hline
		\multicolumn{1}{|l|} {Size of replay memory} & 100000              \\\hline
		\multicolumn{1}{|l|} {Size of mini-batch} & 64                     \\\hline
		\multicolumn{1}{|l|} {Target network update interval} & 1000 steps \\\hline
		\multicolumn{1}{|l|} {Loss function} &  Mean squared error         \\\hline
		\multicolumn{1}{|l|} {Optimizer} & Adam                            \\\hline 
		\multicolumn{1}{|l|} {Activation function} & ReLu                  \\\hline
        
	\end{tabular}
\end{table}

\subsection{DDQN training results}
To assess the training performance of the proposed multi-agent DRL approach, we observe the cumulative rewards per training episode and the behavior of the loss function during the training process.

\begin{figure}[ht]
	\includegraphics[width=\linewidth]{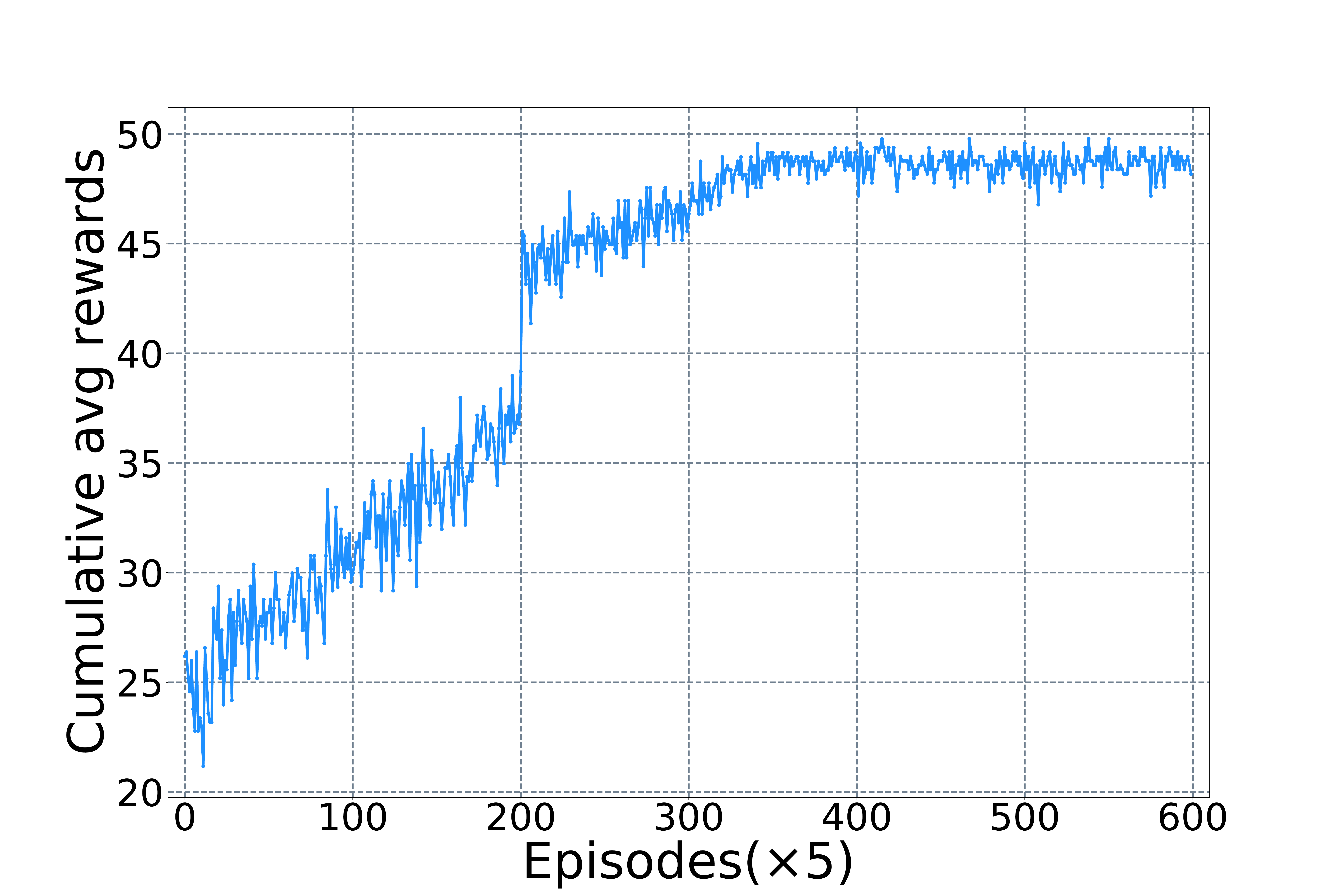}
	\caption{Training rewards.}
	\label{fig:3}
\end{figure}

Fig.~\ref{fig:3} shows the cumulative average rewards of agents per episode. From this figure, the cumulative rewards improve as the number of training episodes increases, which demonstrates the effectiveness of the proposed training algorithm. When the training episode approximatively reaches 1800, the agents gain interesting experiences and start to exploit better actions. Accordingly, the cumulative reward approaches to a maximum value, indicating that the training process converges after an acceptable number of training episodes. Note that the convergence of the DQL algorithm does not present large fluctuations which are principally due to the low-mobility of the end-users in the environment.

\begin{figure}[ht]
	\includegraphics[width=\linewidth]{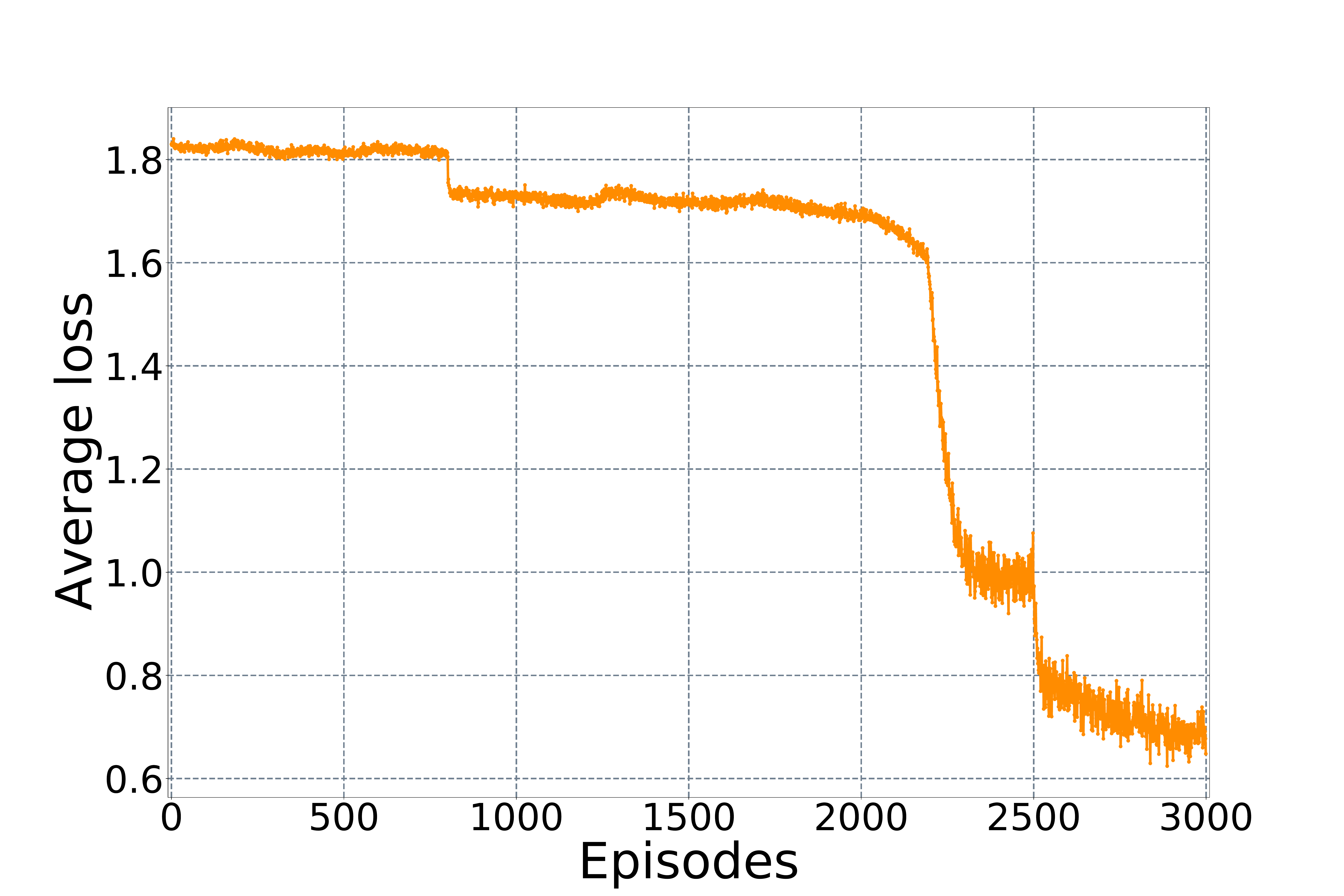}
	\caption{Training loss.}
	\label{fig:4}
\end{figure}

Fig.~\ref{fig:4} illustrates the convergence of the loss function, Eq.~\eqref{loss}, during the training process of the DDQNs. It plots the average of the agents’ loss results versus the episodes. The loss decreases with the increase in training episodes. In the first few episodes, the loss declines gradually because various new actions were explored randomly and as learning progressed, good actions were selectively performed based on the Q-value function that had become more reliable. Consequently, after episode 2500, the loss converges to a minimum value, which demonstrates the accurate Q-value approximation.

 It is worth noting that the convergence results of the reward and loss function are obtained by assuming that the gNodeBs have perfect knowledge of the channel state information, which is used as input of the DDQN algorithm. In the case of imperfect channel state information, the convergence time of the DDQN algorithm may increase \cite{kaur2020imperfect}. Therefore, the DDQN algorithm takes more time to accurately learn the appropriate policies. But, once the offline training of the DDQN algorithm is performed, the learned policy can be applied rapidly to obtain the resource allocation solution.

\subsection{SAMA-RL performance evaluation}
 To benchmark our SAMA-RL approach, we implemented two schemes from the literature, which we called 3-SRA \cite{rw2} and 1-SRA \cite{ofdma} for three-stage resource allocation and one-stage resource allocation, respectively.  3-SRA allocates radio resources to end-users by considering two types of slices, namely the rate constrained slice (i.e., eMBB slice) and the delay constrained slice (i.e., URLLC slice). In the first stage, a central controller uses a heuristic algorithm to reserve radio resources (i.e., a fraction of the system bandwidth) to the eMBB and URLLC slices in each gNodeB. In the second stage, the reserved radio resources for each slice in each gNodeB are adjusted using a DRL algorithm. In the final stage, a heuristic algorithm is deployed to map the fraction of bandwidth assigned to a slice with the physical RBs. Accordingly, we have chosen 3-SRA since it: (i) is a competitive approach that performs the slicing operation of radio resources in a hierarchical manner, (ii) is integrated into a network architecture similar to our proposed architecture, for instance, the presence of a central controller that has a global view of the RAN, (iii) considers two types of slices, namely the eMBB slice and the URLLC slice, and (iv) is based on a DRL algorithm to solve the radio resource allocation problem in the RAN. 

 1-SRA performs the allocation of RBs to end-users in a single-stage process using a greedy heuristic algorithm. Each gNodeB uses this algorithm to assign the required RBs to its associated eMBB and URLLC end-users. This algorithm computes the received signal-to-noise ratio (SNR) of the end-users on all the available RBs of a gNodeB. Then, for each end-user, it: (1) identifies the highest SNR given by an RB, (2) calculates the spectral efficiency (SE) that corresponds to the highest SNR, (3) calculates the required number of RBs, which is defined as the ratio of the traffic queue length to the calculated SE, and (4) assigns the required RBs to it. Note that URLLC end-users are prioritized over eMBB end-users, thus the proposed algorithm schedules URLLC end-users first.  We have chosen 1-SRA as a benchmark scheme since: (i) it considers two types of slices, namely the eMBB slice and the URLLC slice, (ii) the objective function is similar to our formulated objective function, which maximizes the total sum data rates of the URLLC and eMBB end-users, and (iii) it considers a minimum data rate for the eMBB end-user and a maximum latency requirement for the URLLC end-users in the problem formulation, which is similar to our optimization problem.

 We adapt 3-SRA and 1-SRA to meet the requirements of our system model. Specifically, in 3-SRA, two gNodeBs cannot serve a single end-user under the 3-SRA algorithm. On the other hand, in 1-SRA, we calculate the achievable data rate of each RB using the Shannon rate formula instead of the modulation and coding scheme and the queue length is given by the packet size of each end-user. To fairly benchmark the performance of SAM-RL against 3-SRA and 1-SRA, most of the simulation parameters are chosen similarly to those used in \cite{rw2} and \cite{ofdma}. Note that the comparison results are averaged over 1000 trials.

\begin{figure}[ht]
	\includegraphics[width=\linewidth]{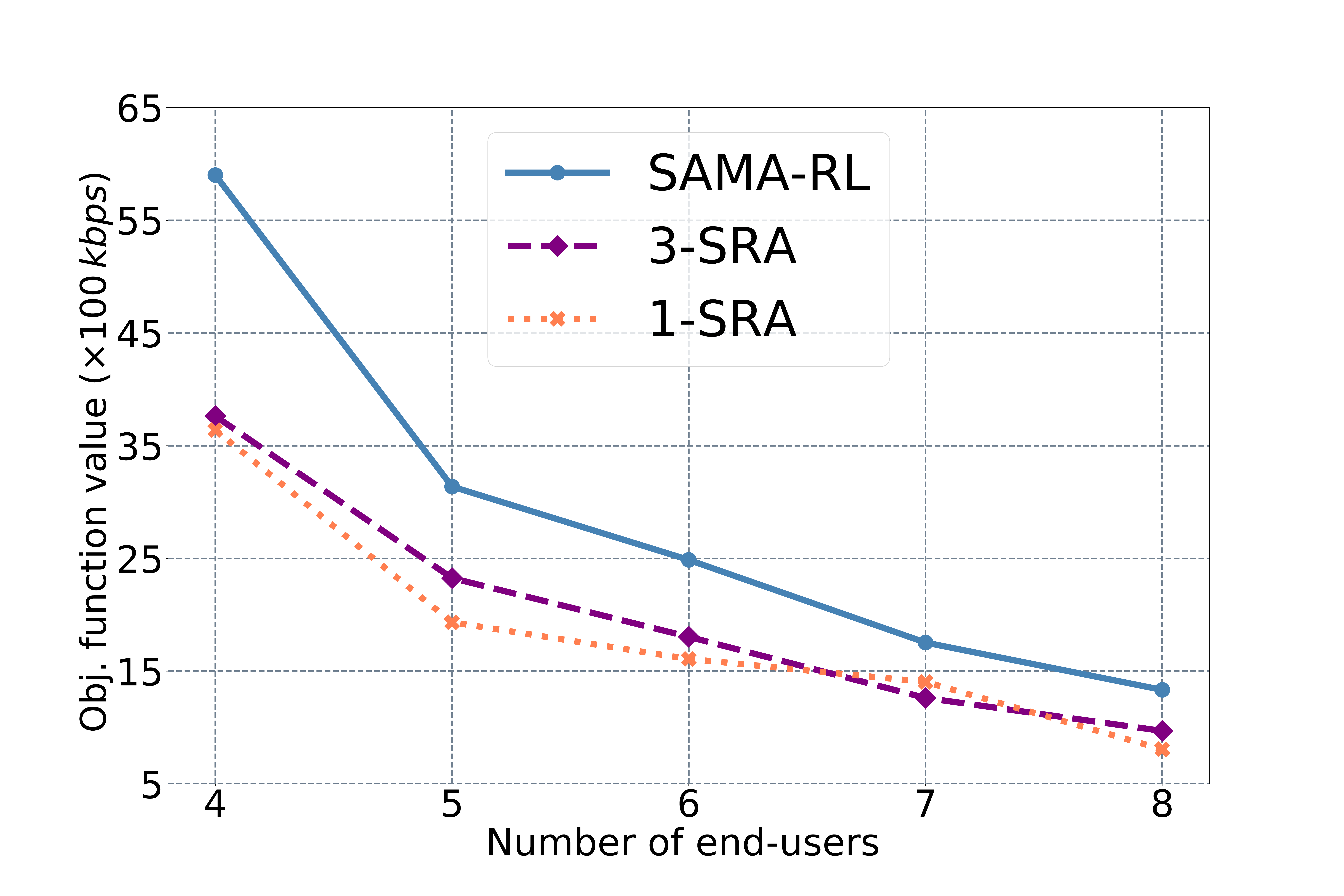}
	\caption{Impact of the number of end-users on the objective function.}
	\label{fig:5}
\end{figure}

Fig.~\ref{fig:5} shows the performance of SAMA-RL compared to the benchmark approaches when varying the number of end-users. SAMA-RL always achieves the best performance in terms of the objective function value when the number of end-users increases. We notice, for the three approaches, that the value of the objective function, which represents the total sum data rates of the URLLC and eMBB end-users, decreases as the number of end-users increases. This is due to the increase of the competitiveness among end-users to obtain a sufficient number of RBs that guarantees their requirements. Indeed, in the case of a low number of end-users, several RBs can be assigned to one end-user to guarantee its requirements in terms of data rate and delay. On the other hand, when the number of end-users is large, each approach seeks to allocate a minimum number of RBs to each end-user in order to satisfy as many end-users as possible in terms of data rate and delay requirements. SAMA-RL maximizes the total sum data rates of the end-users better than 3-SRA since the latter assigns the RBs to the end-users according to the required data rate of the entire slice, which includes a set of end-users, instead of relying on the required data rate of each end-user individually. Therefore, it cannot efficiently allocate the required RBs to the end-users. In the 1-SRA approach, the allocation of RBs is based only on the highest estimated SNR for each end-user, which may poorly determine the RBs required to satisfy the QoS of end-users.

\begin{figure}[ht]
	\includegraphics[width=\linewidth]{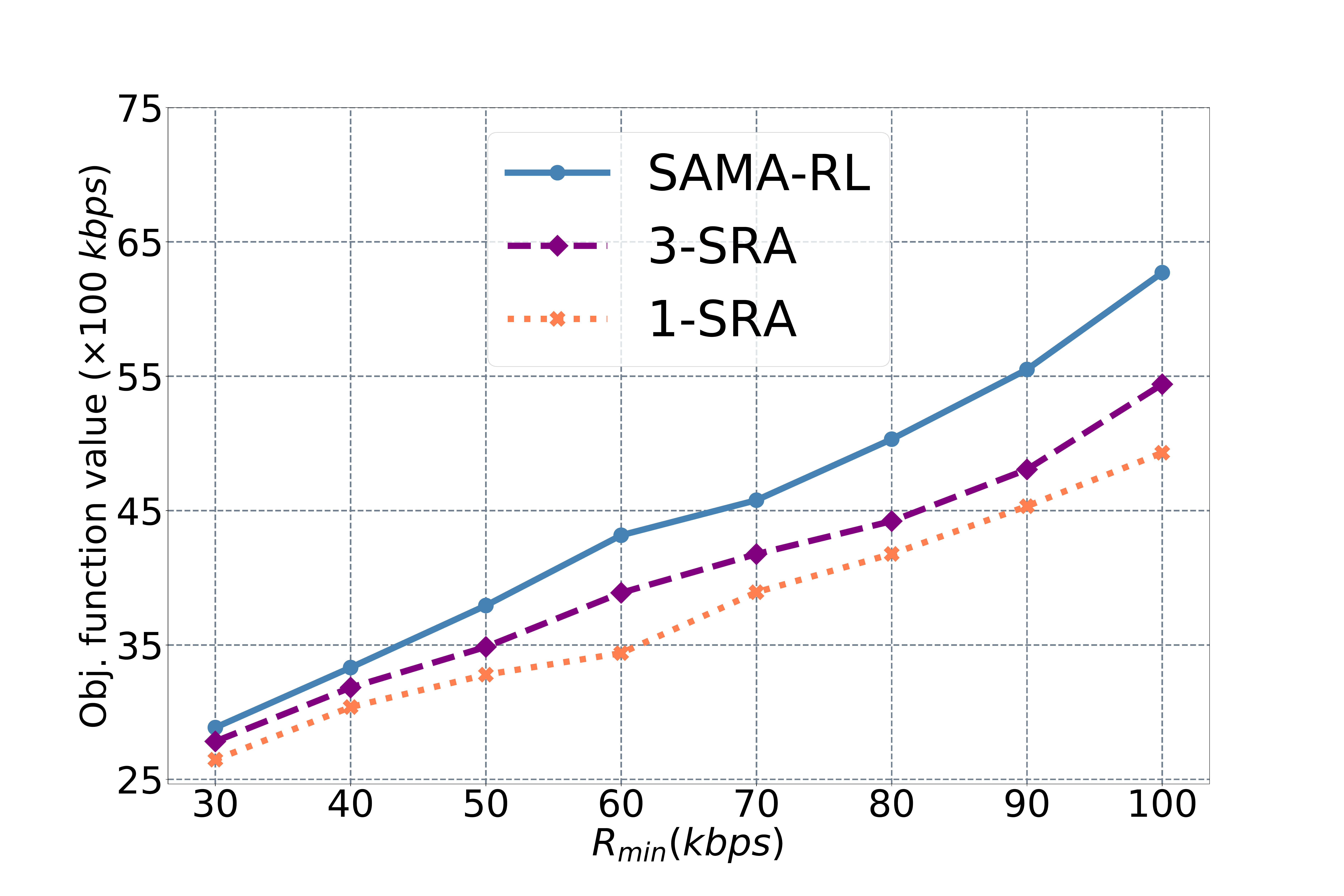}
	\caption{Impact of the minimum data rate threshold ($\mathcal{R}_{min}$) on the objective function.}
	\label{fig:6}
\end{figure}

Fig.~\ref{fig:6} illustrates the impact of the minimum data rate threshold ($\mathcal{R}_{min}$) on the objective function Eq.~\eqref{Obj}. Based on these results, we make the following observations: (1) SAMA-RL outperforms 3-SRA and 1-SRA for all minimum data rate thresholds; (2) the higher is the minimum data rate threshold, the larger is the performance gap between SAMA-RL and both benchmark approaches. This is due to the multi-agent approach proposed in the gNodeB allocation level where a gNodeB can borrow RBs from other adjacent gNodeBs when the RBs pre-allocated by the SDN controller are insufficient. This indeed illustrates the effectiveness of the proposed multi-agent method and demonstrates its robustness.

\begin{figure}[ht]
	\includegraphics[width=\linewidth]{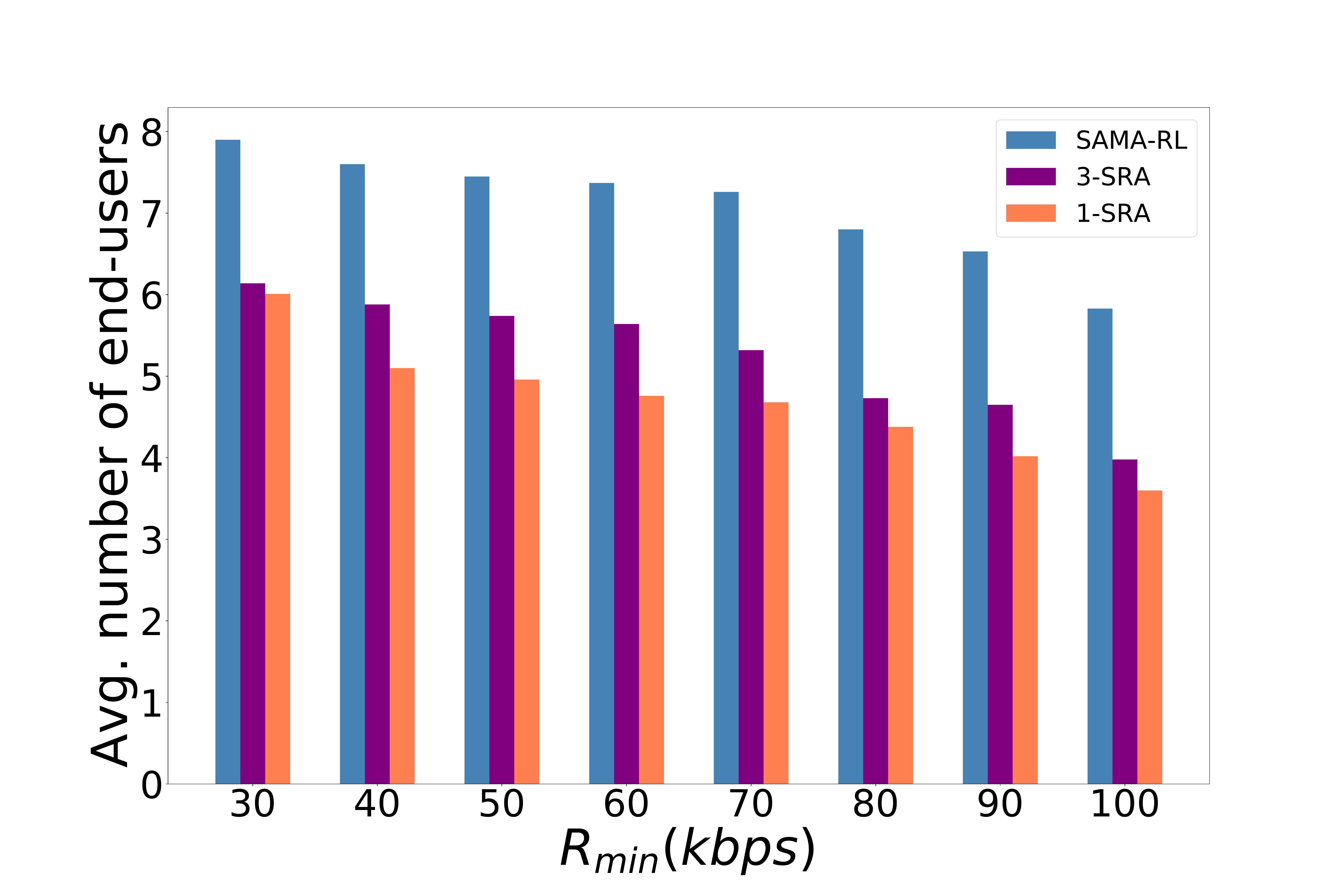}
	\caption{Impact of the minimum data rate threshold ($\mathcal{R}_{min}$) on the number of end-users.}
	\label{fig:7}
\end{figure}

Fig.~\ref{fig:7} presents the comparison of SAMA-RL and both benchmark approaches, i.e., 3-SRA and 1-SRA, in terms of the average number of end-users whose achieved data rate is greater than or equal to the minimum data rate threshold $\mathcal{R}_{min}$. We can see that SAMA-RL always outperforms 3-SRA and 1-SRA for all minimum data rate thresholds. These results can be justified as follows: in 3-SRA, if the reserved resources for a given slice are not sufficient to provide the required service, i.e., minimum data rate threshold, the slice must wait for the next resources reservation update, while in SAMA-RL, a gNodeB can request, if necessary, some RBs from other gNodeB.  In 1-SRA, if the SNR is not appropriately estimated, the RBs allocated to end -users may not be sufficient to provide the required service, thus the end-users will be served in the next resource allocation operation.

\begin{figure}[ht]
	\includegraphics[width=\linewidth]{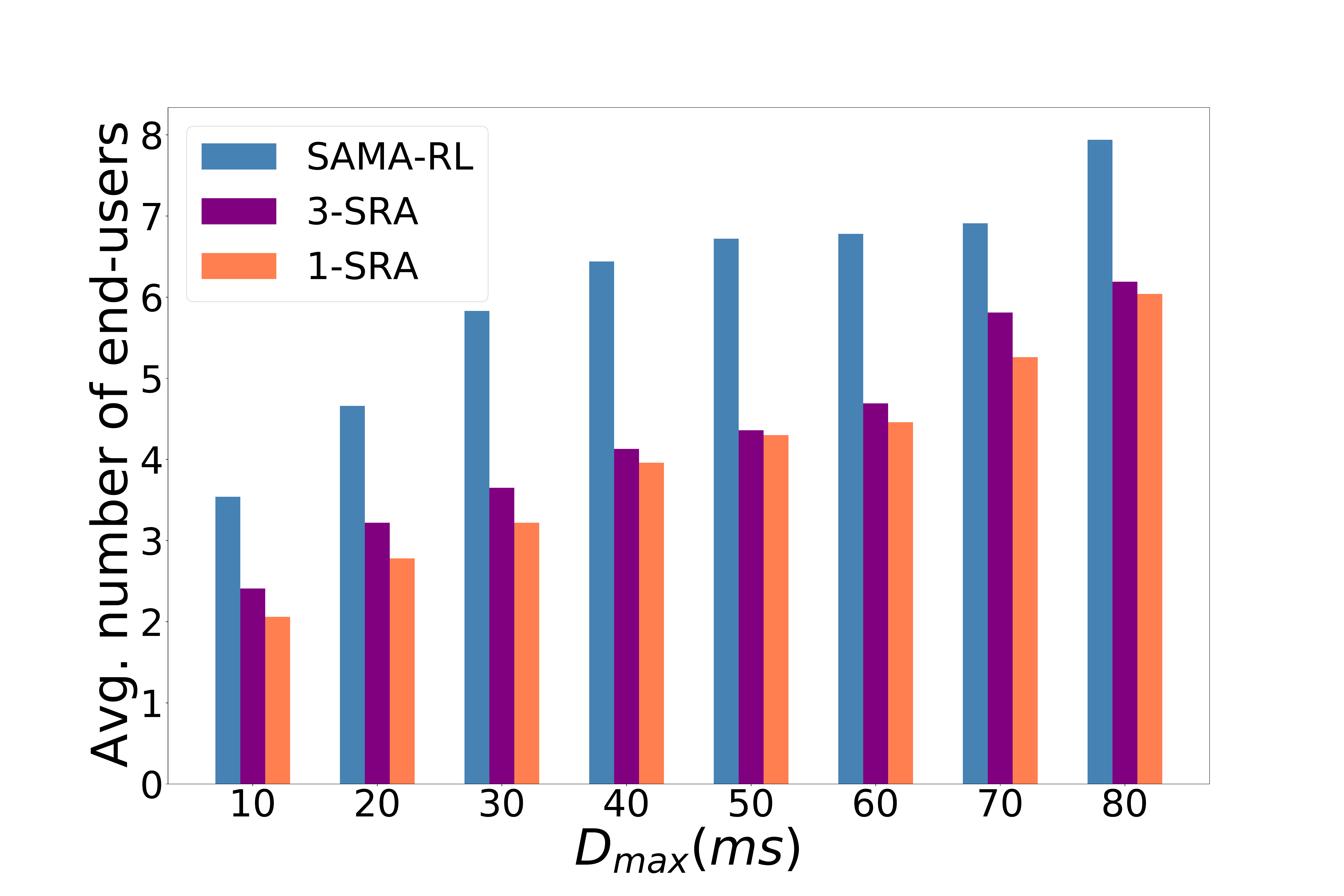}
	\caption{Impact of the the maximum delay threshold ($\mathcal{D}_{max}$) on the number of end-users.}
	\label{fig:8}
\end{figure}

 In Fig.~\ref{fig:8}, we compare the performance of SAMA-RL against the performance both benchmark approaches in terms of the average number of end-users whose delay experienced by a packet is less than or equal to the maximum delay threshold $\mathcal{D}_{max}$. Similar to the obtained results in Fig.~\ref{fig:7}, we can once again confirm that SAMA-RL gives better performance compared to 3-SRA and 1-SRA for all maximum delay thresholds.


\section{Conclusion}
Toward an efficient radio resource slicing of an SDN-enabled RAN, we proposed a two-level RAN slicing mechanism where the allocation of radio RBs is performed in the SDN level and in the gNodeBs level. The SDN level allocates RBs to gNodeBs in a large time-scale while the gNodeBs allocate their RBs to its associated end-users in a short time-scale to efficiently meet their dynamic requirements. Moreover, each gNodeB can borrow some RBs from other gNodeBs when the pre-allocated RBs are insufficient, which decreases the signaling overhead between the two allocation levels  and rapidly provide the needed resources to its end-users. We formulated a data rate maximization problem subject to the ultra-low latency requirements of URLLC services as well as to the minimum data rate requirements of eMBB services. Subsequently, we proved its NP-hardness. Then, we modeled the SDN allocation level problem and the gNodeB allocation problem as a single MDP and a multi-agent MDP, respectively. We adapt the EXP3 algorithm and the DQL algorithm to solve the SDN allocation problem and the gNodeB allocation problem, respectively. In addition, the DQL algorithm applies robust state-of-the-art approaches such as double DQN and replay memory to improve its performance. Simulation results have shown that the proposed two-level mechanism yields significant improvements in terms of RBs allocation. Furthermore, our mechanism outperformed a benchmark algorithm by ensuring the requirements of the URLLC and eMBB services in terms of data rate and delay.

\bibliographystyle{IEEEtran}
\bibliography{IEEEabrv,references}

\end{document}